\documentclass[11pt]{article}
\usepackage[body={16.5cm, 21cm},right=2.8cm]{geometry}
\usepackage{amssymb}
\usepackage{amsmath}

\newcommand\ee{\end{equation}}
\newcommand\be{\begin{equation}}
\newcommand\eea{\end{eqnarray}}
\newcommand\bea{\begin{eqnarray}}

\def\l{\left}
\def\r{\right}
\def\beq{\begin{equation}}
\def\eeq{\end{equation}}

\def\nn{\nonumber}
\def\barr{\begin{array}}
\def\earr{\end{array}}

\def\ie{{\it i.e.}}

\def\D{{\cal D}}

\def\ns{{\nu_5}}
\def\kk{{\kappa_5}}

\begin{document}
\begin{titlepage}
\begin{flushright} 
DFPD-07/TH/09 \\[2.5cm]
\end{flushright}
\begin{center}
{\huge \sc Beyond Twisted Tori} \\[1cm]
{\large  \bf Giovanni Villadoro$^{\, a}$ and Fabio Zwirner$^{\, b}$} \\[.5cm]
{\large \sl $^a$ Jefferson Physical Laboratory, Harvard University, \\
Cambridge, Massachusetts 02138, USA 
\\[.3cm]
$^b$ Dipartimento di Fisica, Universit\`a di Padova and INFN, \\ 
Sezione di Padova, Via Marzolo 8, I-35131 Padova, Italy 
} \\[1.3cm]
{\sc Abstract}
\end{center}
{
Exploiting the fact that Kaluza--Klein monopoles and the associated generalized orbifold   planes are sources for geometrical fluxes, $\omega$, we show that the standard constraint $\omega\,\omega=0$, valid for superstring compactifications on twisted tori, can be consistently relaxed. This leads to novel possibilities for constructing superstring models with fluxes and localized sources, as well as for stabilizing moduli. 
This also explains the ten-dimensional origin of a family of $N=4$ gauged supergravities, whose interpretation in type-IIA orientifold compactifications was lacking.
}

\end{titlepage}


\section{Introduction}

Recent years have witnessed intense theoretical efforts and significant progress in superstring and M-theory compactifications with general systems of fluxes and branes (for recent reviews and references to the original literature, see e.g. \cite{reviews}). However, much remains to be understood in view of a systematic classification of consistent vacua and their low-energy effective field theories, and of a systematic phenomenological analysis of those vacua most closely resembling the observed world.

A rich spectrum of possibilities is offered, already at the classical level (\ie, before the inclusion of perturbative and non-perturbative corrections), by toroidal type-II orientifold compactifications, where we can simultaneously consider fluxes for the RR $p$-forms $G^{(p)}$ and for the NSNS 3-form $H$, as well as geometrical fluxes $\omega$ \`a la Scherk--Schwarz \cite{SS}. The latter modify the topology of the internal manifold, which for this reason is also called `twisted torus' in the literature. 
Recently, this type of fluxes has received a renewed interest in string compactifications for several reasons, such as the possibility to construct new gauged supergravities~\cite{newgaugings}, stabilize moduli~\cite{DKPZ,VZIIA,CFI}, improve the understanding of string dualities~\cite{KSTT,VZD} and consistency conditions~\cite{VZD,M,BK,LB}.
In the presence of O-planes and D-branes, acting as localized sources, a number of stringent constraints must be satisfied, which can be interpreted as integrability conditions associated with the Bianchi identities (BI) of the different local symmetries. Until now, geometrical fluxes were restricted to obey the consistency condition~\cite{SS}
\beq
\omega \, \omega = 0 \, , \nn
\eeq
together with the integrability conditions coming from the BI of the RR and NSNS forms, involving the other fluxes and branes. In this letter we will show that the above condition is too restrictive. Using the fact that 5+1-dimensional Kaluza--Klein monopoles (KK5) and the associated generalized orbifold planes (KKO5) are sources for geometrical fluxes, we will indeed show that, in the presence of these KK sources, the BI for geometrical fluxes modifies into
\beq
d\omega+\omega \, \omega=Q_{KK} \, ,  \nn
\eeq
where $Q_{KK}$ stands for the contribution of the KK sources. Using string dualities, we will also show that the last condition is equivalent to the well-known BI for the NSNS and RR forms. 

The modified BI above tells us that the integrability condition $\omega\,\omega=0$ can be relaxed by adding KK sources. This is analogous (actually dual) to what happens in the RR sector, where the non-trivial RR plus NSNS flux contribution to the BI is cancelled by the contribution of  D-branes and O-planes. The consistency of such constructions is also strengthened by the existence \cite{DKPZ} of consistent $N=1$ truncations of $N=4$ gauged supergravities, derived from type-IIA, $N=1$ orientifold compactifications with NSNS, RR and geometrical fluxes, which do not satisfy \cite{VZIIA} the quadratic constraint~$\omega\,\omega=0$. The inclusion of KK sources allows to fill the gap and to understand the ten-dimensional origin of such gauged supergravities.

Finally, we will conclude by discussing the importance that KK5 monopoles may have in various aspects of string compactifications. 


\section{From KK5~monopoles to generalized twisted tori}

It is well known that NS5-branes can be identified as the magnetic sources dual to strings for the NSNS 2-form potential $B$. In compactifications on twisted tori, the corresponding BI  gets modified by a {\em torsion} term $\omega$,
\beq 
\label{eq:HBI}
dH+\omega\, H=[\ns]\,,
\eeq
where $[\ns]$ is the localized 4-form Poincar\'e dual to the NS5-brane world-volume $\nu_5$, $H$ is the 3-form field strength associated with $B$, and (more details on our notation can be found in \cite{VZIIA}):
\beq
\omega H = \frac{1}{4}\, \omega_{ab}^{\ \ e}\, H_{ecd}\; dx^a \wedge dx^b \wedge dx^c \wedge dx^d\,. \nn
\eeq
Here and in the following, we will ignore all numerical and $\alpha^{\, \prime}$ factors related with charges and tensions. The Scherk--Schwarz \cite{SS} parameter\footnote{Our geometrical flux parameter $\omega_{ab}^{\ \ c}$ corresponds to $-f^c_{\ ab}$
in the notation of \cite{SS}.} $\omega$, often called {\em geometrical} (or {\em metric}) flux in the recent literature, corresponds to a background value for the spin connection, and satisfies the condition
\beq
\omega \, \omega = 0 \, , 
\label{eq:ww0}
\eeq
which ensures the closure of the external derivative $(d+\omega)$, modified by the torsion $\omega$, in the new geometry. Requiring the brane to wrap a non-trivial cycle on the twisted torus corresponds to the constraint \cite{VZD,M}
\beq
(d+\omega) \, [\ns]=0\,, \nn
\eeq
\ie\ $\nu_5$ must be a non-trivial cycle in the cohomology constructed with the modified derivative $(d+\omega)$. The $H$ components in the first term of eq.~(\ref{eq:HBI}) are those sourced by the NS5-brane and by the bulk fluxes (the second term on the l.h.s.). They are not constant on the internal manifold, but have a singularity at the brane, according to Gauss law, in order to satisfy the BI and the equations of motion locally. However, since fields are periodic over the cycles of the original torus, these terms do not contribute to the integrability conditions. On the other hand, non-vanishing background fluxes for $H$ contribute to the BI of eq.~(\ref{eq:HBI}) with the extra torsion term $\omega H$: such term may give a non-trivial contribution to the integrability conditions, which must be compensated by the contributions from localized sources $[\ns]$, as shown in \cite{KM} for the heterotic case\footnote{In \cite{KM} the non-trivial $\omega H$ term was actually compensated by a topological instanton number Tr($F\wedge F$) from the bulk gauge sector: its contribution, however, is globally equivalent to that of a stack of NS5-branes.}. This means that on twisted tori the total charge from NS5-branes can be non-zero, as long as it is eventually cancelled by fluxes, similarly to what happens for D-branes in the presence of RR and NSNS fluxes (see e.g. \cite{GKP}). 

There exist also magnetic sources for $H$ with negative charge and tension. From the effective field theory point of view, they are the NSNS analogue of what the orientifold planes are for the RR forms: they are non-dynamical codimension-four objects with a $Z_2$ orbifold involution under which bulk fields have non-trivial internal parities. For instance, their existence can be deduced in the type-IIB theory by S-duality. There is a string-string duality \cite{HT,DK,W,Sen1} connecting the SO(32) heterotic theory on $T^4$ with the type-IIA theory on K3. As an intermediate step~(see Table~\ref{tab:ss}), we get a type-IIB string theory with 32 D5-branes (16 D5-branes plus their orientifold images), two on each of the 16 O5-planes lying at the fixed points of $T^4/Z_2$. The D5-branes provide the matter fields dual to the U(1)$^{16}$ sector of the heterotic theory, while the O5-planes are there to cancel RR tadpoles and to halve the number of supersymmetries. Via S-duality, the D5-branes map into NS5-branes, which now carry NSNS charge. To cancel this charge, the O5-planes must map into some generalized orbifold planes (NSO5) that carry negative tension and NSNS charge, and act non-trivially on the fields (\cite{K}, see also \cite{P}). Then we must also include these contributions in eq.~(\ref{eq:HBI}), which now reads
\begin{table}
\begin{center}
\begin{tabular}{c|ccccccccc}
&heterotic & $\xrightarrow{\ S\ }$ & type-I & $\xrightarrow{\ T_4\ }$ & type-IIB & $\xrightarrow{\ S\ }$ &  type-IIB & $\xrightarrow{\ T_1\ }$ &type-IIA \\ \hline
on & $T^4$ &&  $T^4$ && $T^4/Z'_2$ && $T^4/Z''_2$ && $T^4/Z_2$ \\
with & U(1)$^{16}$ && D9/O9 && D5/O5 && NS5/NSO5 && KK5/KKO5 
\end{tabular}
\end{center}
\caption{\small \it String-string duality chain between the heterotic theory on $T^4$ and the type-IIA theory on $T^4/Z_2$. The symbols ``$\xrightarrow{\ S\ }$" and ``$\xrightarrow{\ T_n\ }$" mean S-duality or T-duality along $n$ directions (inside $T^4$); $Z'_2$, $Z''_2$ and $Z_2$ are the $Z_2$ involutions of the O5 orientifold, of the NSO5 orbifold and of the KKO5 orbifold, respectively; the last line describes the system of branes/solitons providing the 16 $N=4$ vector multiplets.
\label{tab:ss}}
\end{table}
\beq \label{eq:NSBI}
dH+\omega\,H=Q_{H}\,,
\eeq
where 
\beq 
\label{eq:QNS}
Q_{H}=\sum \Bigl ( [\nu_5]+[\nu^o_5] \Bigr)
\eeq
is the sum of all the contributions from NS5-branes and NSO5-planes ($[\nu^o_5]$). Notice that the latter  give a negative contribution to eq.~(\ref{eq:QNS}), but we reabsorbed the negative charge coefficient in the definition of $[\nu^o_5]$. 

In the absence of fluxes, the integrability condition from eq.~(\ref{eq:NSBI}) implies that the number of NS5-branes must be 32, to cancel the contributions from the NSO5-planes. This condition is just the dual of the RR-tadpole cancellation condition that ensures the cancellation of anomalies. In more general compactifications, however, with $\omega$ and $H$ fluxes, the contributions from NS5-branes need not match the ones from NSO5-planes, analogously to what happens in the RR sector in the presence of RR and NSNS fluxes.

As the NSNS 2-form, also the graviton possesses its own magnetic source in ten dimensions: the Kaluza--Klein 5D monopole \cite{HT,T}.  Its geometry is described by the Euclidean 4D Taub--NUT metric embedded in 10D space-time \cite{TaubNUT,S,GP,T} 
\beq 
\label{eq:KK5metric}
ds^2_{\kk}=\eta_{\mu\nu} dx^\mu dx^\nu + f^{-1}(r)\bigl(dr^2+r^2d\theta^2+r^2\sin^2\theta d\phi^2\bigr)+f(r)\bigl(d\psi+V^\psi \bigr)^2\,,
\eeq
where the indices $\mu,\nu$ span 5+1 space-time dimensions, and
\beq
f(r)=\l(1+\frac{m}{r}\r)^{-1}\,, \qquad V^\psi=m(1-\cos \theta)d\phi\,.  \nn
\eeq
The above metric can be derived from the NS5-brane soliton background via T-duality~\cite{BFRM,OV} (see also \cite{BJO,Tong}), which indeed rotates the metric and the $B$-field.  Eq.~(\ref{eq:KK5metric})  is a solitonic solution of the 10D equations of motion that sources a flux for the graviphoton $V^{\psi}$. This means that KK5~monopoles are the sources for the geometrical fluxes~\cite{KSTT}
\beq
\omega^{\psi}\equiv dV^\psi=m\sin\theta\, d\theta \wedge d\phi\,,
\eeq
which thus satisfy the `BI-like' condition
\beq 
\label{eq:wBI}
d\omega^{\psi}=[\kk]^\psi\,,
\eeq 
where $[\kk]^\psi$ is the localized 3-form in the ($r,\theta,\phi$) space  dual to the world-volume of the KK5 monopole. The components of $\omega$ sourced by the KK5 monopoles are not constant---they must be singular at the monopole---but they are periodic over the torus cycles, therefore the l.h.s. of eq.~(\ref{eq:wBI}) will vanish if integrated over a torus cycle. This means that the total KK charge must also vanish on the torus. Indeed, in analogy with what  we recalled for the BI of $H$, also $\omega$ possesses magnetic sources with negative charge, one example is provided by the Atiyah-Hitchin spaces \cite{AH}. They are solitonic (everywhere smooth) solutions of the 10D Einstein equations, which at large distances look like, up to exponentially small corrections (see also \cite{GM}), KK5 monopoles  with negative charge (and tension), modded by an orbifold involution acting on the four orthogonal directions. Therefore eq.~(\ref{eq:wBI}) will receive also contributions from these generalized orbifold planes ($\kappa_5^o$), namely
\beq 
\label{eq:wBIO}
d\omega^{\psi} = \sum \Bigl ( [\kk]^\psi+[\kappa_5^{o}]^\psi\Bigr) \,.
\eeq
From  the effective field theory point of view, these new objects can be considered just as orbifold fixed planes, which however carry negative tension and source a negative geometrical flux. The integrability condition from eq.~(\ref{eq:wBI}) ensures that we must dress each of these KK orbifold planes (KKO5-planes) with KK5 monopoles, to cancel the negative charge. The localized fields arising from each stack of KK5 monopoles provide the twisted sector of the orbifold. When all KK5 monopoles are put on top of the KKO5-planes, so that tensions and charges cancel locally, the internal manifold is flat, it is just a toroidal orbifold. The twisted sector is provided by the fields localized on KK5~monopoles, and the twisted cycles are those created by the KK5 with their orbifold images, which are indeed shrunk to zero size. An example of these configurations can be obtained by T-dualizing the type-IIB  NS5-brane/NSO5-plane configuration discussed above, which gives type-IIA on the $T^4/Z_2$ orbifold limit of K3, \ie\ the outcome of the string-string duality chain mentioned before. On each of the 16 orbifold fixed points there are a KKO5-plane and two KK5 monopoles (one KK5 plus its image), giving a shrunk 2-sphere and a massless U(1) vector multiplet. The vector field, which can be seen as the SO(2) truncation of the enhanced U(2) gauge group associated to the two KK5~\cite{Sen2}, comes from the RR 3-form over the shrunk sphere.  When the KK5 monopoles move away from the KKO5-planes, the localized fields describing the position of the KK5 monopoles acquire a VEV, the twisted cycles blow up and the internal manifold becomes a K3 at a generic point of its moduli space. The latter can thus be seen as a generic configuration of KK5 and KKO5: indeed, the moduli space and the topology of K3 coincide with those of the KK system describing $T^4/Z_2$ \cite{Seiberg,AM}.

Now, if we compactify on twisted tori, we have to take into account also the contributions from background geometrical fluxes, which turn on other components of $\omega$ than those appearing in eqs.~(\ref{eq:wBI}) and (\ref{eq:wBIO}). Thus we expect that also eq.~(\ref{eq:wBI}) and (\ref{eq:wBIO}) get modified by a torsion term, analogously to the BI for $H$ of eq.~(\ref{eq:NSBI}). We can derive the modified BI by applying Buscher rules~\cite{B}  directly to eq.~(\ref{eq:HBI}). Indeed, under T-duality along one direction, say $c$,  the components of the NSNS 3-form flux along $c$ map into a geometrical flux, a NS5-brane orthogonal to $c$ goes into a KK5 monopole, with the fibered $S_1$ along the dualized direction, and analogously NSO5-planes map into KKO5-planes, namely
\begin{eqnarray}
H_{abc} & \longrightarrow & \omega_{ab}^{\ \ c} \, , \nn \\
\left[ \nu_5 \right]_{dabc} & \longrightarrow & [ \kappa_5 ]_{dab}^{\ \ \ \ c} \, , \nn \\
\left[ \nu_5^o \right]_{dabc} & \longrightarrow & [ \kappa_5^o ]_{dab}^{\ \ \ \ c} \, . \nn
\end{eqnarray}
Hence eq.~(\ref{eq:NSBI}) goes into\footnote{If  $\omega$ is along the T-dualized direction, \ie\ $\omega_{bc}^{\ \ \ d}$, it T-dualizes into a 'non-geometric' flux~\cite{KSTT}: we will restrict ourselves here to geometric compactifications, but we will comment on the non-geometric case below.} 
\beq
d\omega+\omega \, \omega = Q_{KK} \, , 
\label{eq:wwBI}
\eeq
where
\beq
Q_{KK} = \sum \Bigl ( [\kk]^\psi+[\kappa_5^{o}]^\psi\Bigr)  \, .
\label{eq:QKK}
\eeq
We thus get also a non-trivial contribution from the background geometrical fluxes of the twisted tori, which now contribute non-trivially to the integrability conditions (\ref{eq:wwBI}). This means that it is possible to have a non-vanishing total KK charge in a compact space, as long as $\omega \, \omega\neq 0$.  This seems to be in contrast with the consistency condition of eq.~(\ref{eq:ww0}). However, the topology in the presence of a KK5 monopole changes---a string wrapping the fibered circle $S_1$ may unwrap passing through the tip of the monopole~\cite{GHM}. Eq.~(\ref{eq:wwBI}) suggests that, when (and only when) eq.~(\ref{eq:wwBI}) is satisfied, this change exactly compensates for the apparent clash with the closure of the external derivative. Having separately either a violation of eq.~(\ref{eq:ww0}) or an uncancelled net KK charge in a compact volume would lead to a topologically inconsistent construction. However, when both are present and satisfy the integrability condition from eq.~(\ref{eq:wwBI}), the whole construction is consistent. Exactly as it works for the T-dual construction with NS5-branes and $H$ fluxes on twisted tori discussed before. 

Another cross-check comes from `S-duality' in type-IIA. In the strong coupling limit, the type-IIA theory uplifts to M-theory, developing a new dimension. KK5~monopoles in type-IIA derive from KK6~monopoles in M-theory, when one of the worldvolume dimensions of the KK6 is compactified to a circle that shrinks to zero size \cite{BJO}. Analogously, KKO5-planes uplift to M-theory by adding a dimension in their worldvolume. Therefore, eq.~(\ref{eq:wwBI}) uplifted to M-theory should keep the same form,
\beq
d\omega+\omega \, \omega=Q_{KK}\,, \nn
\eeq
with the only difference that now the space is eleven-dimensional. We can now recover type-IIA in another limit, by shrinking a different circle to zero. If we identify the eleventh dimension with $\psi$, the fibered $S_1$ of the KK6, the latter will produce a D6-brane in ten dimensions \cite{T}; Atiyah-Hitchin spaces will give O6-planes \cite{SW} (see also \cite{Sen2}); $\omega^\psi$ will instead map into the RR 2-form flux $G^{(2)}$~(see e.g.~\cite{KSTT,DP}) since the RR 1-form $C^{(1)}$ is given by the graviphoton $V^\psi$. In this limit eq.~(\ref{eq:wwBI}) will then reduce to the type-IIA equation
\beq 
\label{eq:G2BI}
dG^{(2)}+\omega \, G^{(2)}=Q_{RR}=\sum \Bigl ( [\pi_6]+[\pi^o_6]\Bigr) \,,
\eeq 
which is the BI for the RR form  $G^{(2)}$ in the presence of D6-branes ($\pi_6$) and O6-planes ($\pi_6^o$) on twisted tori \cite{VZIIA}.  Eqs.~(\ref{eq:wwBI}) and (\ref{eq:G2BI}) have thus the same M-theory origin, confirming the consistency of compactifications with non-vanishing global KK charge, cancelled by geometrical fluxes with non-trivial  $\omega \, \omega$.

Notice that in principle we could saturate the contributions from O6-planes by using fluxes instead of D6-branes, so that among the light degrees of freedom there is no extra localized matter field besides the bulk sector. Analogously, we could saturate the negative contributions from the KKO5-planes with geometrical fluxes $\omega$ and no KK5~monopoles. In this case the compactification would keep the orbifold involution, but without light twisted sectors! These compactifications thus provide deformations of the orbifold that stabilize/avoid the twisted fields. Consider, for instance, the family of $AdS_4$ $N=1$ supersymmetric vacua from the superpotential of refs.~\cite{DKPZ,VZIIA} in the type-IIA  theory compactified on the $T^6/(Z_2\times Z_2)$ orbifold, with O6-planes, generic fluxes and all untwisted bulk moduli stabilized; by saturating now the RR and KK BI just with fluxes, without D6-branes and KK5~monopoles, {\em all} moduli are stabilized, since in this case there are no extra light fields from the orbifold twisted sector nor from D-branes. 
  
We can also derive the intrinsic O6-orientifold parity of the KK5~monopoles by using M-theory. As mentioned before, the M-theory uplift of an O6-plane is an Atiyah-Hitchin space, which at large distances can be well approximated by a KK6~monopole solution with negative mass parameter $m$,  with the 11th dimension identified with the $S_1$ of the monopole ($\psi$), and with a $Z_2$ involution on the four orthogonal dimensions.  The KK5~monopole of the type-IIA theory comes from another KK6 in M-theory, this time extending along the $\psi$ direction. Since this direction is odd under the $Z_2$, the KK5~monopole worldvolume is odd under the O6-orientifold involution. This is analogous to what happens to D4-branes, which come from M5-branes wrapping the 11th dimension. This means that the KK5~monopole can only wrap 2-cycles that are odd under the O6 parity. In usual $N=1$ compactifications on orbifolds or CY with O6-planes, all 2-cycles are odd with respect to the orientifold involution. Therefore, unlike NS5-branes, KK5~monopoles can be consistently included in $N=1$ orientifold string compactifications to four dimensions. In particular, this also means that, in the study of the effective action for the bulk moduli, in $N=1$ type-II compactifications with generic fluxes and branes, condition (\ref{eq:ww0}) can be relaxed by inserting KK5 monopoles. 

Finally, notice that eq.~(\ref{eq:wwBI}) nicely fits with the T-duality invariant form  for the NSNS Bianchi identities
(up to non-geometrical fluxes), namely
\beq
\D\D=Q_{NS}\,, \label{eq:DBI}
\eeq
where $\D$ is the modified external derivative in the presence of geometrical and $H$ fluxes,
\beq
\D=d+\omega+H\wedge \ ,
\eeq
$Q_{NS}$ is the sum of the NSNS sources
\beq
Q_{NS}=Q_{H}+Q_{KK}\,, \nn
\eeq
and eq.~(\ref{eq:DBI}) should be read as projected into a basis of forms. In particular, eq.~(\ref{eq:DBI}) gives back eqs.~(\ref{eq:HBI}) and (\ref{eq:wwBI}). 

In the case of non-geometrical fluxes we can guess that eq.~(\ref{eq:DBI}) will continue to hold, with $\D$ replaced by the combination $d+H+\omega+Q+R$ (where $Q$ and $R$ are the non-geometrical fluxes T-dual to $H$ and $\omega$ defined in \cite{STW}), and with $Q_{NS}$ receiving also contributions from the solitons (if they exist) sourcing $Q$ and $R$ fluxes. 


\section{An example: the DKPZ solution}

In ref.~\cite{DKPZ}, Derendinger, Kounnas, Petropoulos and one of the authors (DKPZ) derived the effective $N=1$ superpotential for bulk moduli in  a $T^6/(Z_2\times Z_2)$ type-IIA compactification with generic NSNS, RR and geometrical fluxes,  exploiting the underlying $N=4$ supergravity that is present after the orientifold projection, but before the $Z_2 \times Z_2$ orbifold projection. In particular, they found an $AdS_4$ solution with exact $N=1$ supersymmetry and all closed untwisted moduli stabilized. Later \cite{VZIIA} it was realized that, although such vacua correspond to a consistent $N=4$ gauged supergravity from the effective field theory point of view, they do not admit an interpretation in terms of geometric compactifications from ten dimensions with fluxes, D-branes and O-planes. This is due to the fact that, in contrast with the heterotic theory \cite{KM}, there does not seem to be a one-to-one correspondence between BI constraints from the compactification  and Jacobi identities of the underlying $N=4$ gauged supergravity. While the fact that some of the compactifications cannot be viewed as $N=4$ gaugings can be easily understood, in terms of $N=1$ D-brane configurations that realize part of the $N=4$ supersymmetry in a non-linear way, the $N=4$ gaugings without an interpretation in terms of compactifications from ten dimensions were not understood. In particular, the DKPZ $AdS_4$ vacua fail to satisfy eq.~(\ref{eq:ww0}). We can now understand why it was not possible to obtain these vacua from compactifications with only fluxes and D6/O6 sources: they also need KK sources.  In the notation of \cite{VZIIA}, eq.~(\ref{eq:ww0}) for the DKPZ setup reduces to the conditions
\bea 
\label{eq:wwDKPZ}
\omega_3(\omega_3-\omega_1)&=&0\,, \nn \\
\omega_2(\omega_3-\omega_1)&=&0\,, 
\eea
where $\omega_{1,2,3}$ correspond to some components of the geometrical fluxes. The first condition corresponds to a Jacobi identity of the underlying $N=4$ gauged supergravity, which however does not require the second condition. The DKPZ $AdS_4$ vacua have $\omega_3=0$  but non vanishing $\omega_1$ and $\omega_2$. This means that the second condition is not satisfied, thus the compactification requires the existence of a mismatch between the charges of KK5~monopoles and KKO5-planes. In particular, it is easy to check that the number of required KK5 monopoles is less than the one needed to cancel the charge and the tension from the KKO5-planes, \ie\ the needed KK charge and tension are negative. In some sense, we need to remove part of the KK5 monopoles present at the $T^6/(Z_2\times Z_2)$ fixed points. Each of them must wrap one of the three factorized 2-tori (which are indeed odd under the orientifold involution), with fibered $S_1$ parallel to the O6-plane. The fact that the vacuum solution in the effective 4D theory is supersymmetric, and agrees with the constraints from $N=4$ supergravity, tells us that such KK5 monopole configurations preserve the same $N=4$ supersymmetry as the O6-plane, as for D6-branes parallel to the O6-plane.

It is also possible to check that, once the right number of KK5 monopole is removed, the effective potential derived by dimensional reduction (the contribution of the KK5 monopoles is discussed in \cite{LE}) agrees with the one dictated by supersymmetry and by the $N=1$ superpotential of \cite{DKPZ,VZIIA}, once (and only once) eq.~(\ref{eq:wwBI}) is satisfied.  In particular, the contributions to the effective potential for the closed string moduli coming from the tensions of KK5 monopoles and KKO5-planes cancel against the extra contributions from the Einstein term, originating from the non-closure of the external derivative ($\omega\, \omega \neq 0$). KK sources and the possibility of violating eq.~(\ref{eq:ww0}) thus fill the gap in the understanding of the 10D origin of the DKPZ type-IIA vacua and of the microscopic interpretation of the corresponding $N=4$ gaugings. 

Notice that, as in the case of D6-branes, there also exist consistent KK5-monopole configurations  that do not correspond to any $N=4$ gauging, but preserve at least $N=1$ supersymmetry in four dimensions. They correspond to KK5 monopoles with the $S_1$ fiber orthogonal to the O6-planes, giving a non-vanishing contribution also to the condition in the first line of eq.~(\ref{eq:wwDKPZ}). When such KK5~monopoles are inserted (or removed) the corresponding vacua cannot be seen as a truncation of an $N=4$ gauged supergravity, analogously to what happens when D6-branes at generic angles are considered, as discussed in \cite{VZIIA}.


\section{Outlook}

As mentioned before, by performing suitable Scherk--Schwarz twists, it is possible to add KK5 monopoles to $N=1$ type-IIA compactifications with intersecting branes and fluxes. Besides relaxing the condition of eq.~(\ref{eq:ww0}), and allowing for new vacua with geometrical fluxes, these new objects contribute to the effective action with extra matter  fields localized on the KK5 monopoles. As in the case of D6-branes, each KK5 monopole gives an $N=4$ vector multiplet in four dimensions (a non-Abelian group is generated if a stack of KK5 is considered~\cite{Sen2}), eventually truncated by orbifold and orientifold projections and with mass terms from flux contributions. These extra matter fields can also be seen as arising  from the bulk RR $p$-forms, calculated  on the new cycles generated by the KK5 geometry, analogously to what happens in type-IIA on the $T^4/Z_2$ orbifold mentioned before, with fixed points resolved by KK5~monopoles and KKO5-planes. 

There is a number of interesting aspects of these new compactifications that would be worth studying. First of all, they may add new phenomenologically relevant ingredients to the usual models with intersecting/magnetized branes, both because they allow to relax the condition (\ref{eq:ww0}) and because KK5 monopoles can generate extra matter fields at low energy. Moreover, KK5~monopoles might be a new source of SUSY breaking: changing their orientation with respect to other localized objects (such as O-planes and D-branes) and/or turning on localized magnetic fluxes we expect that D~terms may be generated, analogously to the case of usual D-branes. From a complementary point of view, it would be interesting to understand better the embedding of these new compactifications in the general classifications of gauged supergravities (see e.g.~\cite{SW2,max,e11}) and of generalized geometries for superstring compactifications (see e.g.~\cite{GMPT}). 

We focused here on type-II compactifications,  but we could also consider heterotic and type-I theories on these new geometries. Since these string theories give at most $N=4$ supergravities in four dimensions and KK5~monopoles are BPS, thus breaking half of the supersymmetries, a configuration with non-trivial KK charge, eventually cancelled by $\omega$ fluxes, cannot be viewed as a $N=4$ gauging in 4D, unlike type-II compactifications where BPS objects can be included in an $N=4$ invariant way. This is the reason why in the heterotic and type-I theories the correspondence between BI and Jacobi identities of the $N=4$ gaugings is one-to-one, while this is not the case in type-II or M-theory: with $N=1$ supersymmetry in ten dimensions, the only $N=4$ supergravities in four dimensions arise from compactifications without localized sources. This, however, does not forbid the construction of $N=2,1,0$ vacua in heterotic/type-I compactifications with non-trivial KK monopole charges. Finally, we did not discuss much the extension to non-geometrical fluxes (see e.g.~\cite{non-geo,STW,VZD}); it would be interesting to understand whether there exist analogous sources for non-geometrical fluxes too, and if they may allow to relax the corresponding BI constraints~\cite{STW}. We leave all this to future work.

\section*{Acknowledgments}
We thank C.~Angelantonj, M.~Bianchi, G.~Dall'Agata, C.~Kounnas and M.~Porrati for discussions and comments on the manuscript. We would like to thank the organizers of the workshop {\em Superstring Phenomenology}  and the Kavli Institute for Theoretical Physics in Santa Barbara for their kind hospitality. GV also thanks the Galileo Galilei Institute for Theoretical Physics for the hospitality and the INFN for partial support during the completion of this work. This research was supported in part by the European Programme ``The Quest For Unification'', contract MRTN-CT-2004-503369, and by the U.S. National Science Foundation under Grant No. PHY99-07949.

\end{document}